\documentclass[reqno]{amsart}
\usepackage{amssymb}
\usepackage{amsmath}

\usepackage{appendix}

\usepackage{hyperref}

\usepackage{color}

\newcommand{\eq}[1]{\eqref{#1}}
\newcommand\beq{\begin{equation}}
\newcommand\eeq{\end{equation}}

\newtheorem{theorem}{Theorem}[section]
\newtheorem{proposition}[theorem]{Proposition}
\newtheorem{lemma}[theorem]{Lemma}
\newtheorem{corollary}[theorem]{Corollary}

\DeclareMathOperator{\supp}{supp}
\DeclareMathOperator{\tr}{tr}

\numberwithin{equation}{section}

\renewcommand\H{\mathcal{H}}
\renewcommand\L{\mathrm{L}}
\newcommand\R{\mathbb R}
\newcommand\N{\mathbb N}

\newcommand\Z{\mathbb Z}

\newcommand\D{\mathcal{D}}

\newcommand\I{\mathcal{I}}

\newcommand\di{\mathrm d}

\newcommand{\T}{\mathcal{T}}

\newcommand\e{\mathrm{e}}
\newcommand\eps{\varepsilon}
\newcommand\vphi{\varphi}

\newcommand{\la}{\langle}
\newcommand{\ra}{\rangle}

\renewcommand\P{\mathbb P}
\newcommand\E{\mathbb E}

\newcommand{\bom}{{\boldsymbol{{\omega}}}}

\newcommand{\qnorm}[1]{\left\lVert\!\left\Vert #1 \right\rVert\!\right\rVert}
\newcommand{\abs}[1]{\left\lvert #1 \right\rvert}
\newcommand{\norm}[1]{\left\lVert #1 \right\rVert}
\newcommand{\scal}[1]{\la #1 \ra}

\newcommand{\set}[1]{\left\{ #1 \right\}}
\newcommand{\pa}[1]{\left( #1 \right)}

\begin{document}

\author[F. Germinet]{Fran\c cois Germinet}
\address[Germinet]{Universit\'e de Cergy-Pontoise,
CNRS UMR 8088, IUF, D\'epartement de Math\'ematiques,
F-95000 Cergy-Pontoise, France}
\email{germinet@math.u-cergy.fr}

\author[A. Klein]{Abel Klein}
\address[Klein]{University of California, Irvine,
Department of Mathematics,
Irvine, CA 92697-3875,  USA}
\email{aklein@uci.edu}

\author[B. Mandy]{Benoit Mandy}
\address[Mandy]{Universit\'e de Cergy-Pontoise,
CNRS UMR 8088, D\'epartement de Math\'ematiques,
F-95000 Cergy-Pontoise, France}
\email{mandy@math.u-cergy.fr}


\title[Delocalization for unbounded random Landau Hamiltonians]{Delocalization for random Landau Hamiltonians with unbounded random variables}

\begin{abstract}
In this note we prove the existence of a localization/delocalization transition for Landau Hamiltonians randomly perturbed by an electric potential with unbounded amplitude. In particular, with probability one, no Landau gaps survive as the random potential is turned on{; the gaps close, filling up partly with localized states.}  A minimal rate of transport is exhibited in the region of delocalization. To do so, we exploit the a priori quantization of the Hall conductance and extend recent Wegner estimates to the case of unbounded random variables.
\end{abstract}

\maketitle

\section{Introduction}
\label{sectintro}

In this note we prove the existence of a dynamical localization/delocalization transition for Landau Hamiltonian randomly perturbed by an electric potential with unbounded amplitude, extending results from  \cite{GKS1,GKS2}. In \cite{GKS1} the perturbation had to be sufficiently small compared to the strength of the magnetic field: the amplitude of the random potential was  such that  the Landau gaps survived after adding the perturbation. In \cite{GKS2} the Landau gaps where allowed to close, but the  random  potentials were  bounded. In this article we consider random potentials such that, with probability one,   all  the Landau gaps close as the random potential is turned on, and are shown to be (partially) filled up with localized states. As in \cite{GKS1,GKS2}, a minimal rate of transport is exhibited in the region of delocalization.

{ These results} exploit the a priori quantization of the Hall conductance proved in \cite{GKS2}.  Many of the results we will need rely on \cite{GK1,GK5}, where the random potential was assumed to be bounded.  Such a strong assumption is not necessary, and can be replaced by  weaker hypotheses,  satisfied by the random Landau Hamiltonian with unbounded random couplings studied in this paper. We will require
Wegner estimates  for these random operators,  which are obtained by
extending the analysis of \cite{CHK1,CHK2} to the case of unbounded random variables, a result of  independent interest.

We now describe the model and the results.
We consider a
$\mathbb{Z}^2$-ergodic Landau Hamiltonian
\begin{equation} \label{landauh}
H_{B,\lambda,\bom} =H_B +
\lambda  V_{\bom} \quad \mathrm{on} \quad
\mathrm{L}^2(\mathbb{R}^2, {\mathrm{d}}x),
\end{equation}
where $H_B$ is the (free) Landau Hamiltonian,
\begin{equation}\label{HB}
H_B =  (-i\nabla-\mathbf{A})^2 \quad \text{with} \quad
\mathbf{A}= \tfrac B2 (x_2,-x_1).
\end{equation}
($\mathbf{A}$ is the vector potential and
$B>0$ is the strength of the magnetic field, we use the symmetric gauge
and incorporated the charge of the electron in the vector potential),
$\lambda \ge 0$ is
the disorder parameter,   and
$V_\omega$ is an unbounded ergodic potential:  there is a probability space $(\Omega, \P)$ equipped with an ergodic group $\{\tau(a); \ a \in \Z^2\}$ of
measure preserving transformations, a potential-valued map $V_\omega$ on $\Omega$, measurable in the sense that $\langle \phi, V_\bom \phi \rangle$ is a measurable function of $\omega$ for all $\phi \in \mathcal{C}^\infty_c(\mathbb{R}^2)$.  We  assume that
\begin{equation}\label{defV}
V_\bom(x) =\sum_{j\in\Z^2} \omega_j u(x-j),
\end{equation}
where  the single site potential $u$ is a  nonnegative bounded
measurable function
on $\R^{d}$ with compact support, uniformly
bounded away from zero in
a neighborhood of the origin,  and  the $\omega_j$'s are  independent, identically distributed  random variables,  whose common  probability distribution $\mu$ has  a bounded density $\rho$  with  $\supp \rho= \R$ and fast decay:
\beq \label{rhodecay}
\rho(\omega) \le \rho_0 \exp(-|\omega|^\alpha),
\eeq
for some $\rho_0\in]0,+\infty[$ and $\alpha>0$.   We fix constants for $u$ by
\begin{equation}
C_{-}\chi_{\Lambda_{\delta_{-}}(0)}\le u \le C_+ \chi_{\Lambda_{\delta_{+}}(0)}\quad \text{with $C_{\pm}, \delta_{\pm}\in ]0,\infty[ $},
\end{equation}
and normalize $u$ so that we have  $\|\sum_{j\in \Z^2} u_j\|_\infty\le 1$.  (We write
$
\Lambda_{L}(x):= x +\left[-\tfrac L 2, \tfrac L 2\right[^d
$
for the
box of side $L>0$ centered at $x\in \R^2$, with $\chi_{\Lambda_{L}(x)} $ being its characteristic function.  We also write $\chi_x=\chi_{\Lambda_{1}(x)} $.)

Under these hypotheses,  $H_{B,\lambda,\bom}$ is  essentially self-adjoint on $\mathcal{C}_c^\infty(\R^d)$ with probability one,
with the bound  {  $\pa{ \scal{x}:=\sqrt{1 +\abs{x}^2}}$}
\begin{equation}
H_{B,\lambda,\bom}\ge - c_\omega (\log \scal{x})^\beta, \mbox{ for all } x\in\R^d,
\end{equation}
for any given $\beta>\alpha^{-1}$, { with} $c_\omega$ depending also on $\alpha,\beta,d$. {  (See Lemma~\ref{lembb}.)}

{ Moreover, the unbounded random potential $V_\omega$ satisfies
the probability estimate of  Lemma~\ref{lembb}, namely  \eq{controlVL},    the condition that replaces the boundedness of the potential in \cite{GK1,GK5}.  Note that  \eq{controlVL} is similar to the condition given in \cite[Eq.~(3.2)]{U}.   Using the Wegner estimate given in Theorem~\ref{thmWegner}, we can conclude, similarly to the results in  \cite{U} for a continuous Gaussian random potential, that the results of \cite{GK1,GK5}, and hence also \cite{GK3,GKjsp}, hold for $H_{B,\lambda,\bom}$.  (See also Appendix~\ref{sectmsa}.)
This condition also suffices for the validity of \cite[Theorems~1.1 and 1.2]{GKS2}.
Thus we just refer to \cite{GK1,GK3,GK5,GKjsp,GKS2} where appropriate.}

The spectrum $\sigma(H_B)$  of the  Landau
Hamiltonian $H_B$ consists of a sequence of infinitely
degenerate eigenvalues, the
Landau levels:  {
\begin{equation} \label{landaulevels}
B_n=(2n-1)B ,\quad n=1,2,\dotsc .
\end{equation}}
For further reference, we also set  {
\beq
\mathcal{B}_1=]-\infty,2B[,\quad \mathrm{and} \quad \mathcal{B}_n=]B_n-B,B_n+B[, \quad n=2,3,\dotsc .
\eeq}
On the other hand, as soon as $\lambda>0$, the spectrum fills the Landau gaps and we have \cite{BCH}
\beq
\sigma(H_{B,\lambda,\bom})=\R, \quad \P-\mbox{a.s.}
\eeq
The fact that the Landau gaps are immediately filled up as soon as the disorder is turned on implies that  the approach used in \cite{GKS1} is  non applicable. More properties of the Hall conductance are needed in order to perform the simple reasonning that provides the existence of a dynamical transition. More precisely, it becomes crucial to know a priori that the Hall conductance is an integer in the region of complete localization (which includes the spectral  gaps), a fact that was circumvented in \cite{GKS1} by resorting to an open gap condition. That the Hall conductance for ergodic models is integer valued in the localization region was known for discrete Anderson type models since \cite{BES,AG}. For ergodic Schr\"odinger operators in the continuum, it was first established in \cite{ASS} for energies in gaps and extended to the region of complete localization in \cite{GKS2}, where the analysis of \cite{AG} has been carried over to the continuum. This property has to be combined with the continuity of the Hall conductance for arbitrary small $\lambda$ (in order to let $\lambda$ go to zero). In \cite{GKS2} it is shown that it is actually enough to prove the same continuity property but for the integrated density of states; see \cite[Lemma~3.1]{GKS2}. This is done in this note by revisiting the article \cite{HKS}; see Theorem~\ref{holderc}. But first, we extend the Wegner estimate given in \cite{CHK2} to unbounded random variables;  the estimate is given in terms of  the concentration function  of a measure  which is a modification  of the single-site probability measure $\mu$. (See Theorem~\ref{thmWegner}, which has independent interest.)

We state the main result of this note and its corollary. Following \cite{GK5,GKjsp,GKS1,GKS2}, we set $\Xi_{B,\lambda}^{\mathrm{DL}}$ to be the region of complete localization (gaps included), that is, the set of energies where  the multiscale analysis applies (or,  if applicable,   the fractional moment method of \cite{AENSS}). Its complement  is the set of dynamical delocalization $\Xi_{B,\lambda}^{\mathrm{DD}}$. An energy $E\in\Xi_{B,\lambda}^{\mathrm{DD}}$ such that for any $\eps>0$, $[E-\eps,E+\eps]\cap\Xi_{B,\lambda}^{\mathrm{DL}}\not=\emptyset$, is called a dynamical mobility edge.

\begin{theorem}\label{thmlimit2}  { Let { $H_{B,\lambda,\bom}$} be a  random Landau Hamiltonian}   as above.
For
each $n=1,2,\dots$, if $\lambda$ is small enough (depending on $n$) there exist dynamical mobility edges $\widetilde{E}_{j,n}(B,\lambda)\in\mathcal{B}_n$,
$j=1,2$, such that
\begin{align}
\max_{j=1,2} \left \lvert  \widetilde{E}_{j,n}(B,\lambda)  - B_n  \right \rvert 
\le  K_n(B)\lambda
\abs{\log \lambda}^{\frac 1 \alpha} \to  0 \quad
\text{as $\lambda \to 0$},
\label{lambda20}
\end{align}
with a finite  constant  $ K_n(B) $.
(It is possible
that  $\widetilde{E}_{1,n}(B,\lambda)=
\widetilde{E}_{2,n}(B,\lambda)$, i.e., dynamical delocalization occurs at
a single energy.)
\end{theorem}

By the characterization of the region of complete localization established in \cite{GK5}, Theorem~\ref{thmlimit2} has a consequence in terms of transport properties of the Hall system.
Indeed, to measure ``dynamical delocalization" as stated in the theorem, we introduce
\begin{equation}\label{moment}
M_{B,\lambda,\omega}(p,\mathcal{X},t)  =
\left\|  {\langle} x {\rangle}^{\frac p 2}
{\mathrm{e}^{-i tH_{B,\lambda,\omega} }}
\mathcal{X}(H_{B,\lambda,\omega}) \chi_0
\right\|_2^2 ,
\end{equation}
the
random moment of order
$p\ge 0$ at time $t$ for the time evolution  in the Hilbert-Schmidt norm, 
initially spatially localized in the square of side one around the origin
(with characteristic function $\chi_0$), and ``localized"
in energy by the function $\mathcal{X}\in C^\infty_{c,+} (\mathbb{R})$.
Its time averaged expectation is given by
\begin{equation}   \label{tam}
\mathcal{M}_{B,\lambda}( p ,\mathcal{X}, T )   = 
\frac1{T} \int_0^{\infty}
\mathbb{E}\left\{ M_{B,\lambda,\omega}(p,\mathcal{X},t)\right\}
{\mathrm{e}^{-\frac{t}{T}}} \,{\rm d}t .
\end{equation}

\begin{corollary}\label{cordeloc}
The random Landau Hamiltonian 
$H_{B,\lambda,\omega}$
exhibits   dynamical delocalization
in each Landau band $\mathcal{B}_n(B,\lambda)$:
For each  { $n=1,2,\ldots$}  there exists at least one energy
$E_n(B,\lambda)\in \mathcal{B}_n(B,\lambda) $,
such that for every 
$\mathcal{X}\in \mathcal{C}^\infty_{c,+} (\mathbb{R})$  with
$\mathcal{X} \equiv 1$  on some open interval  $J\ni E_n(B,\lambda) $
and  $p>0$,  we have
\begin{equation}\label{momentgrowth}
\mathcal{M}_{B,\lambda}(p,\mathcal{X}, T)    \ge \
C_{p,\mathcal{X}} \, T^{\frac p4 - 6} \ ,
\end{equation}
for all  $T \ge 0$  with  $  C_{p,\mathcal{X}} > 0 $.
\end{corollary}

As mentioned aboved, to prove Theorem~\ref{thmlimit2} we extend  the  Wegner estimate of  \cite{CHK2} to measures $\mu$ with unbounded support. More precisely, the finite volume operator $H_\omega^{(\Lambda)}$ satisfies  extensions of the Wegner estimates of  \cite{CH,CHK1,CHK2}.  As in \cite{CHK2},  we do not require the probability measure $\mu$ to have a density. Precise statements and proofs are given in Appendix~\ref{appWegner}.

\section{Hall conductance and dynamical delocalization}
\label{sectproof}

We start by introducing some notation.  Given $p \in [1,\infty)$,
$\T_p$ will denote  the Banach space of bounded operators $S$
on $\mathrm{L}^2(\mathbb{R}^2, {\mathrm{d}}x)$ with
$\| S \|_{\T_p}=\| S \|_p \equiv\left(\tr |S|^p\right)^{\frac 1p} < \infty$.
A random operator $S_\omega$ is a strongly measurable
map from the probability
space $(\Omega,\P)$ to bounded operators on
$\mathrm{L}^2(\mathbb{R}^2, {\mathrm{d}}x)$.   Given $p \in [1,\infty)$,
we set
\begin{equation}
\qnorm{S_\omega}_p\equiv
\left\{ \E \left\{  \| S_\omega \|_p^p   \right\}\right\}^{\frac 1p}=
\left\lVert  \| S_\omega \|_{\T_p} \right\rVert_{\text{L}^p(\Omega,\P) },
\end{equation}
and
\begin{equation}
\qnorm{S_\omega}_\infty\equiv
\left\lVert  \| S_\omega \| \right\rVert_{\text{L}^\infty(\Omega,\P) }.
\end{equation}

We define the  $(B,\lambda, E)$ parameter set by
\[
\Xi=
\left\{(0,\infty)\times [0,\infty) \times \R\right\}\backslash
\cup_{B \in (0,\infty)} \{(B,0)\times \sigma(H_B) \}  ;
\]
that is we exclude the Landau levels at no disorder.  We set
$$
P_{B,\lambda,E,\omega}=\chi_{]-\infty,E]}(H_{B,\lambda,\omega}).
$$
The Hall conductance $\sigma_H(B,\lambda,E)$ is given by (e.g.\cite{BES,ASS,AG,BGKS,GKS1,GKS2})
\begin{equation}\label{sigmaH}
\sigma_H (B,\lambda,E) = - 2\pi i\,
\E \left\{\tr \left\{\chi_0 P_{B,\lambda,E,\omega}
\left[ \left[P_{B,\lambda,E,\omega},X_1\right],
\left[P_{B,\lambda,E,\omega},X_2\right]\right]
\chi_0\right\}\right\},
\end{equation}
defined for $(B,\lambda, E) \in\Xi$ such that
\begin{equation} \label{welldef}
\qnorm{\chi_0 P_{B,\lambda,E,\omega}
\left[\left[ P_{B,\lambda,E,\omega},X_1\right],
\left[ P_{B,\lambda,E,\omega},X_2\right]\right] \chi_0}_1 < \infty.
\end{equation}
($X_i$ denotes the operator given by multiplication by the coordinate
$x_i$, $i=1,2$, and  $|X|$ the operator given by multiplication by $|x|$.) In particular, $\sigma_H (B,\lambda,E)$ is well-defined for all $(B,\lambda,E)$ such that $E\in \Xi_{B,\lambda}^{\mathrm{DL}}$. Moreover it is proved in \cite{GKS2} that $\sigma_H (B,\lambda,E)$ is integer valued for all $(B,\lambda,E)$ such that $E\in \Xi_{B,\lambda}^{\mathrm{DL}}$. We need to investigate the continuity properties of
$\sigma_H (B,\lambda,E)$, as $\lambda$ tends to zero.
In \cite{GKS2} we prove that for any $(B,\lambda,E)$ such that $E\in \Xi_{B,\lambda}^{\mathrm{DL}}$, for any $p>1$, there exists a constant $C(p,B,\lambda,E)<\infty$ for any
$(B',\lambda',E')$ in a neighborhood of $(B,\lambda,E)$,
\begin{align}
& |\sigma_H(B',\lambda',E')-\sigma_H(B,\lambda,E)|  \label{sigmadif} \\
& \quad \le C(p,B,\lambda,E)\sup_{u \in \Z^2}\qnorm{\chi_0  \left(P_{B',\lambda',E',\omega} -P_{B,\lambda,E,\omega}\right) \chi_u}_1^{\frac 1p} .\notag
\end{align}
We shall combine this fact with the following proposition, a consequence from  Theorem~\ref{holderc}, which includes an extension of \cite{HKS} to unbounded random variables.

\begin{proposition}
Let $I$ be an open interval in a spectral gap of $H_B$. Then for all $\lambda \ge 0$ the Hall conductance is H\"older continuous in $E\in I$, and   for any $E\in I$ the Hall conductance at Fermi energy $E$ is  H\"older continuous in the disorder parameter $\lambda\ge 0$. 
\end{proposition}

\begin{proof} The proposition is a direct consequence of Theorem~\ref{holderc} and \eq{sigmadif}.
\end{proof}

\begin{proof}[Proof of Theorem~\ref{thmlimit2}]
We set
\begin{equation}\label{LB}
L_B= K_B \sqrt{\tfrac{4\pi}  B } , \quad
\N_B  = L_B \N, \quad \text{and} \quad   \Z^2_B  = L_B \Z^2.
\end{equation}
Note that $L_B\ge 1$ may not be an integer. We consider squares
$\Lambda_L(0)$ with $L \in \N_B$ and identify them with the torii
$\mathbb{T}_L:=\R^2/(L\Z^2)$
in the usual way.  We further let $\widetilde{\Lambda}_L(x)= \Z^2 \cap \Lambda_L(x)$. Given
$L \in {{\N_B}}$ we define finite volume Landau Hamiltonians $ H_{B,0,L} $ on $ \mathrm{L}^2(\Lambda_L(0))$ as in \cite[Section~5]{GKS1},   and set
\begin{equation}\begin{split}
\label{landauh2}
H_{B,\lambda,0,L,\omega}& = H_{B,0,L} +
\lambda  V_{0,L,\omega} \quad \mathrm{on} \quad
\mathrm{L}^2(\Lambda_L(0)),\\
V_{0,L,\omega}(x)&=  \sum_{i\in  \widetilde{\Lambda}_{L-\delta_u}(0)}
\omega_i \,u(x-i),
\end{split}\end{equation}
It follows from  \eq{rhodecay} that
\begin{equation}
\mu(\{ \abs{u} \ge \eps\})\le
C_\alpha \exp\left(-\tfrac12 |\eps|^\alpha\right)\quad \text{for all}\quad \eps >0.
\end{equation}
Let ${\bar{L}} \in \N_B$ (see \eqref{LB}), and let $H_{B,\lambda,0,{\bar{L}},\omega}$
and $V_{0,{\bar{L}},\omega}$ be as in \eqref{landauh2}. A straightforward computation shows that uniformly in $\lambda\in[0,1]$,
\begin{align}\notag
&\P\left\{\sigma(H_{B,\lambda,0,{\bar{L}},\omega})\subset
\bigcup_{n=1}^\infty
[B_n - \lambda\eps, B_n +\lambda\eps]  \right\}\ge
\P\left\{ |\omega_i|\le \eps  \
\text{if ${i\in  \widetilde{\Lambda}_{{\bar{L}}-\delta_u}(0)}$}\right\}\\
\quad & \ge
\left(1 - C_\alpha \exp\left(-\tfrac12 |\eps|^\alpha\right)
\right)^{({\bar{L}}- \delta_u)^2} \ge
1 -C_2 C_\alpha \exp\left(-\tfrac12 |\eps|^\alpha\right) {\bar{L}}^2 \label{nospec}.
\end{align}

We now apply the finite volume criterion for localization given in
\cite[Theorem~2.4]{GK3},  in the same way  as in
\cite[Proof of Theorem~3.1]{GK3},  with parameters (we fix $q \in ]0,1]$)
$\eta_{I,\lambda}=\frac 1 2 \eta_{B,\lambda,I,q} = \frac 1 2 \eta_{B,1,I,q}$ and
$ Q_{B,\lambda,I}\le   \tilde{Q}$,  for some $\tilde{Q}<\infty$ independent of $\lambda\in[0,1]$ as it follows from
Theorem~\ref{thmWegner}. (Note that the fact that
we work with length scales $L \in \N_B$ instead of $L \in 6\N$ only affects
the values of the constants in \cite[Eqs. (2.16) -(2.18)]{GK3}.)

To conduct the multiscale analysis of \cite{GK1,GK3}, we note that in finite volume we have, for any given $\eta<1$, and uniformly in $\lambda\in[0,1]$,
\begin{align}
&\P\left(|\lambda V_\omega(x)| \le L^\eta, \mbox{ for all } x\in\Lambda_L(y)\right)  \\
&\ge
\P\left(|V_\omega(x)| \le  L^\eta, \mbox{ for all } x\in\Lambda_L(y)\right) \\
& \ge 1 - C_\alpha \exp(- \tfrac12  L^{\eta\alpha}) L^2,
\end{align}
which is as close to $1$ as wanted, provided $L$ is large enough (independently of $\lambda$).
Probabilistic bounds on the constant in SLI and EDI follow, with constants bounded by $L^{\eta/2}$. Since we are working in spectral gaps,
we use the Combes-Thomas estimate of    \cite[Proposition 3.2]{BCH} (see also
\cite[Theorem 3.5]{KK1}--its proof, based on \cite[Lemma 3.1]{BCH},
also works for Schr\"odinger operators with magnetic fields),
adapted to finite volume as in  \cite[Section~3]{GK3}.

Now fix $n \in \N$, take $I= \I_n({B})$, and set ${\bar{L}}={\bar{L}}(n,B)$ to
be the smallest
$L \in \N_B$ satisfying \cite[Eq. (2.16)]{GK3}. Let
$E \in \mathcal{I}_n(B), \, \abs{E-B_n} \ge 2 \lambda\eps$,
where $\eps= \eps(n,B,\lambda)) >0$ will be chosen later. Then, using
\eqref{nospec} and the Combes-Thomas estimate, we conclude that 
condition \cite[Eq. (2.17)]{GK3} will be satisfied at energy $E$
if
\begin{gather}
{\eps}\ge C_3\,  ( \log\bar{L})^{\frac 1\alpha},
\\
C_4 \left(\lambda \eps\right)^{-1}{\bar{L}}^{\eta}
\e^{- C_5 \sqrt{\lambda\eps}{\bar{L}}} < 1, \label{condCT}
\end{gather} 
for appropriate constants $C_j=C_j(n,B)$, $j=3,4,5$, with $C_5 > 0$.
This can be done by choosing (in view of \eqref{nospec})
\begin{equation}
\eps = C_3\,  ( \log\bar{L})^{\frac 1\alpha},
\end{equation}
and taking $\bar{L}$ large enough to satisfy \eqref{condCT} depending on
$\lambda \le 1$.
We conclude from \cite[Theorem~2.4]{GK3} that
\begin{equation}\label{smalldisorder}
\left\{  E \in \I_n({B}); \  \abs{E-B_n} \ge C_5 \lambda
\abs{\log \lambda}^{\frac 1\alpha}\right\}\subset 
\Xi_{B,\lambda}^{\text{DL}}.
\end{equation}
for all $\lambda \le 1$.  In particular, for all $n \in \N$ there is $\lambda_n >0$ such that
$B_n-B\in\Xi_{B,\lambda}^{\text{DL}}$ for all $\lambda \in [0,\lambda_n]$.

The existence at small disorder of dynamical mobility edges
$\widetilde{E}_{j,n}(B,\lambda)$, $j=1,2$,
satisfying  \eqref{lambda20}   now follows from \cite{GKS2}
and \eqref{smalldisorder}. Indeed, since  $B_n-B\in\Xi_{B,\lambda}^{\text{DL}}$ for all $\lambda \in [0,\lambda_n]$,  the Hall conductance is constant at energy $B_n-B$ for all $\lambda\in[0,\lambda_n]$. Since for $\lambda=0$, its value is $n - 1$, we can conclude that there is an energy of delocalization between $B_{n}-B$ and $B_n+B=B_{n+1}-B$ for all $\lambda \in [0,\min \set{\lambda_n,\lambda_{n+1}}]$. Then \eqref{smalldisorder} and the constancy of the Hall conductance on sub-intervals of $\Xi_{B,\lambda}^{\text{DL}}$ imply the estimate \eqref{lambda20}.
\end{proof}

\appendix

{\appendixpage}

In these appendices we extend  results known for Anderson-type random Schr\"odinger operator   to unbounded random variables. These appendices are of separate interest and independent of the rest of the paper.

We consider a random Schr\"odinger operator of the form $H_{\lambda,\omega}= H_0 +\lambda V_\omega$  on $ \mathrm{L}^2(\mathbb{R}^d, {\mathrm{d}}x)$, where the random potential $V_\omega$ is as in   \eqref{defV} and $\lambda \ge 0$.
The unperturbed Hamiltonian $H_0$ will be either the Landau Hamiltonian $H_B$ on $\mathrm{L}^2(\mathbb{R}^2, {\mathrm{d}}x)$,  as  in \eq{HB}, or it will have
the general form $H_0= (-i\nabla - A_0)^2 + V_0$ on $ \mathrm{L}^2(\mathbb{R}^d, {\mathrm{d}}x)$, $d \in \N$, where both $A_0$ and $V_0$ are regular enough so that $H_0$ is essentially self-adjoint on $\mathcal{C}_0^\infty(\R^d)$
and bounded  from below by some constant $\Theta\in\R$. As a sufficient condition, it is enough to require that the magnetic potential $A_0$ and the electric potential $V_0$
satisfy the Leinfelder-Simader conditions (cf. \cite{BGKS}):
\begin{itemize}
\item   $A_0(x) \in \mathrm{L}^4_{\mathrm{loc}}(\R^d; \R^d)$  with
$\nabla \cdot A_0(x) \in \mathrm{L}^2_{\mathrm{loc}}(\R^d)$.

\item  $V_0(x)= V_{0,+}(x) - V_{0,-}(x)$ with
$V_{0,\pm}(x) \in \mathrm{L}^2_{\mathrm{loc}}(\R^d)$, $V_{0,\pm}(x) \ge 0$,
and
$ V_{0,-}(x)$  relatively bounded with respect to
$\Delta$ with relative bound $<1$, i.e., there are $0 \le\alpha  < 1$ and $\beta \ge 0$
such that
\begin{equation}\notag
\|V_{0,-}\psi\| \leq
\alpha \| \Delta \psi \| + \beta \|\psi \| \quad \mbox{for all $\psi \in\D(\Delta)$}.
\end{equation}
\end{itemize}
We will say that  $H_0$ is periodic if  $A_0$ and  $V_0$ are $\Z^d$-periodic. It has the property (UCP) if it satisfies the unique continuation principle.  ($H_0$ has the (UCP) if $A_0$ and  $V_0$ are  sufficiently regular; see the discussion in \cite{CHK1}.)

\section{Applicability of the multiscale analysis}
\label{sectmsa}

We provide here estimates that are needed for extending the multiscale analysis, more precisely results of \cite{GK1,GK3,GK5,GKjsp,GKS1,GKS2}, from bounded to unbounded random variables, as mentioned in the introduction. Finite volume operators are as defined in those papers.  We fix the disorder $\lambda \ge 0$ and omit it from the notation. Note that the constants are all uniform in $ \lambda$ for $\lambda \le \lambda_0$.

\begin{lemma}\label{lembb}
Given a box $\Lambda$, there exists $L^\ast$, such that for any $L\ge L^\ast$ we have, for any $\beta>\alpha^{-1}$, {
\beq\label{controlVL}
\P\{\|\chi_{\Lambda_L} V_{\bom}\|_\infty \le   C_+ (\log L)^\beta\} \ge 1 - C(\alpha, \delta_+,d) \rho_0 \exp(- C(\alpha,\beta,\delta_+,d) |\log L|^{\alpha\beta}).
\eeq }
Then for $\P$-a.e.\ $\omega$ we have {
\begin{equation}\label{controlV}
V_{\bom}(x) \ge -c_\omega(\log \scal{x})^\beta \quad \text{for all $x\in\R^d$},
\end{equation}  }
where  $c_\omega>0$ (depending also on $d,\alpha,\beta$). As a consequence
$H_\bom$ satisfies the lower bound
\begin{equation}\label{lbH}
H_\omega \ge - c_\omega (\log \scal{x})^\beta, \mbox{ for all } x\in\R^d,
\end{equation}
for any given $\beta>\alpha^{-1}$ and  is  essentially self-adjoint on $\mathcal{C}_c^\infty(\R^d)$ with probability one.
\end{lemma}

\begin{proof}
To get \eqref{controlVL}, we note that
\beq
\P\{\|\chi_{\Lambda_L} V_{\bom}\|_\infty \le   C_+ (\log L)^\beta\}
\ge
1 - C(2L)^d \P\{|\omega| \ge (\log L)^\beta\}.
\eeq
The bound \eqref{controlV} then follows from the Borel-Cantelli Lemma. Now in view of  \eqref{controlV}, $H_{B,\omega}$ satisfies the lower bound \eqref{lbH} and  thus  $H_{\bom}$ is  essentially self-adjoint on $\mathcal{C}_c^\infty(\R^d)$ with probability one by the Faris-Levine Theorem \cite[Theorem~X.38]{RS2}.
\end{proof}

Bounds on the constant in SLI and EDI follow from \eqref{controlVL}.
GEE follows from heat kernel estimates, as given in \cite{BLM}. As for SGEE, the bound has been derived by Ueki \cite{U} for Gaussian random variables. For the reader's convenience we provide a short proof in the next theorem.  Recall that $H_0\ge \Theta$. We write $E_{H_\omega}(I)=\chi_I(H_\omega)$.

\begin{theorem}\label{thmSGEE}
There exist $m(d)>0$ such that if $\E (|\omega_0|^{m(d)+\alpha}) <\infty$, with $\alpha\ge 0$, then for any bounded interval $I$ we have
\beq
\E\left\{|\omega_0|^\alpha \tr \chi_0 E_{H_\omega}(I) \chi_0\right\} \le C(H_0,d,I,\alpha),
\eeq
for some constant $C(H_0,d,I,\alpha)<\infty$.
Moreover, $m(1)=1$ and $m(d)=2$ for $d=2,3$.
\end{theorem}

\begin{proof}
For simplicity, we assume that the support of $u_0$ is included in the unit cube centered at the origin. If not, straightforward modifications of the argument (as in \cite{CHK2}) yield the result as well.
We write $H=H_\omega=H_0+V_\omega$, with $H_0$ bounded from below, say $H_0\ge 0$. We denote by $E$ the center of the interval $I$. We set $\tilde{I}$ to be the interval $I$ but enlarged by a distance $\tilde{d}:=2|I|$ from above and below: $I\subset \tilde{I}$ and $\mathrm{dist}(I,\tilde{I}^c)=\tilde{d}$. We have
\begin{align}
\tr \chi_0 E_H(I) &= \tr \chi_0 E_H(I)E_{H_0}(\tilde{I}) + \tr \chi_0 E_H(I)E_{H_0}(\tilde{I}^c) \\
& \le C (|E| + 3|I|)^d + \tr \chi_0 E_H(I)E_{H_0}(\tilde{I}^c).
\end{align}
Now, with $R_0(z)=(H_0-z)^{-1}$,
\begin{align}
\tr \chi_0 E_H(I)E_{H_0}(\tilde{I}^c)
&  =
\tr \chi_0 E_H(I)(H_\omega-E-V_\omega) R_0(E) E_{H_0}(\tilde{I}^c) \\
& \le \frac{|I|}{\tilde{d}} \tr \chi_0 E_H(I) \chi_0  + |\tr \chi_0 E_H(I) V_\omega R_0(E) E_{H_0}(\tilde{I}^c )\chi_0|\\
&\le \frac12 \tr \chi_0 E_H(I) \chi_0 + \sum_{j\neq 0} \|\omega_j u_j R_0(E)E_{H_0}(\tilde{I}^c) \chi_0\|_1 \\
& \quad + |\omega_0| |\tr \chi_0 E_H(I) u_0 R_0(E)E_{H_0}(\tilde{I}^c) \chi_0|,
\end{align}
so that, for $p>d$ given, taking advantage of $u_j\chi_0=0$ if $j\neq 0$ (use Helffer-Sj\"ostrand formula plus resolvent identities to get trace class operators),
\begin{align}
\sum_{j\neq 0} \|\omega_j u_j R_0(E)E_{H_0}(\tilde{I}^c) \chi_0\|_1
\le
\E|\omega_0|  \sum_{j\neq 0} C_p (1+|j|)^{-p} .
\end{align}
Next, if $d=1$ then $u_0 R_0(E) E_{H_0}(\tilde{I}^c)$ is trace class, and $\E|\omega_0| < \infty$ is a sufficient condition. If $d=2,3$ (in the present application $d=2$), then Cauchy-Schwartz inequality leads to
\begin{align}
& |\tr \chi_0 E_H(I) \omega_0 u_0 R_0(E)E_{H_0}(\tilde{I}^c)| \\
&\le
\|\chi_0 E_H(I)\|_2 \|\omega_0 R_0(\Theta-1)\chi_0\|_2 \|(H_0+\Theta+1)R_0(E)E_{H_0}(\tilde{I}^c)\|_\infty \\
& \le
\left(1+ \frac{|E|+|\Theta|+1}{\tilde{d}}\right) \|\chi_0 E_H(I)\|_2 \|\omega_0 R_0(\Theta-1)\chi_0\|_2
\\
&\le \frac14 \tr \chi_0 E_H(I) + \left(1+ \frac{|E|+|\Theta|+1}{\tilde{d}}\right)^2 \omega_0^2\tr  \chi_0 R_0(\Theta-1)^2.
\end{align}
The latter trace is finite in dimension $d=2,3$, finishing the proof provided $\E\omega_0^2<\infty$. In higher dimensions, one repeats the very last step as many times as necessary, as in \cite{CHK2}.
\end{proof}

\section{Optimal Wegner estimate with unbounded random variables}
\label{appWegner}

In this appendix we extend  the analyses of \cite{CHK2} and \cite{HKS} to unbounded random variables.

Given a finite box $\Lambda \subset \R^d$, we denote by $H^{(\Lambda)}_{\lambda,\omega}$ an appropriate self-adjoint restriction  of $H_{\lambda,\omega}$ to $\Lambda$, in which case    $H^{(\Lambda)}_{\lambda,\omega}$ has a compact resolvent (see \cite{CHK1,CHK2,GKS1}). There is no other restriction on the boundary condition in Theorem~\ref{thmWegner}(b),(c) below.  When we use the (UCP) for $H_0$ periodic, as in Theorem~\ref{thmWegner}(a), we assume  periodic boundary condition as in \cite{CHK2}.  If $H_0=H_B$, the Landau Hamiltonian, in Theorem~\ref{thmWegner}(a) we assume finite volume operators as defined in \cite[Section~4]{GKS1}  and used in \cite[Section~4]{CHK2}.

If $\Delta$ is a Borelian, $E_{H^{(\Lambda)}_{\lambda,\omega}}(\Delta)$ denotes the associated spectral projection for $H^{(\Lambda)}_{\lambda,\omega}$.

In this appendix we assume $0\le \lambda\le 1$ since we are mostly interested in small values of the coupling constant, but arguments easily extend to $\lambda\le \lambda_0$ for any given $\lambda_0$.

Given an arbitrary Borel measure $\nu$ on the real line, we set  $Q_\nu(s)$ to be a multiple of its concentration function:
\beq
Q_\nu(s):=
8 \sup_{a \in \R}
\nu([a,a+s])
\eeq
Note that $Q_\nu(s)< \infty$ if $\nu$ is a finite measure. The Wegner estimate in \cite{CHK2} is stated in terms of $Q_\mu$; in our extension to unbounded measures $Q_\mu$ is replaced by  $Q_{\mu^{(q)}}$, for an appropriate $q\ge 1$, where
$d\mu^{(q)}(s):=|s|^qd\mu(s)$ for $q>0$.

\begin{theorem}\label{thmWegner}
Consider $H_{\lambda,\omega}$  with $0< \lambda\le 1$. There  exists $1\le m(d)<\infty$,  such that if $\E \{ |\omega_0|^{m(d)}\}<\infty$,   given $E_0\in \R$:
\begin{itemize}
\item[(a)]
Assume either $H_0=H_B$ or $H_0$ is periodic with the (UCP).  Then there exists a  constant $K_W(\lambda)$, depending also on $d$, $E_0$, $\delta_{\pm}$ and $C_{\pm}$, such that for  any compact
interval $\Delta\subset ]-\infty,E_0[$ we have
\beq\label{Wegner}
\E \set{\tr E_{H^{(\Lambda)}_{\lambda,\omega}}(\Delta)} \le K_W(\lambda) Q_{\mu^{(m(d))}}(|\Delta|)|\Lambda|.
\eeq
\item[(b)] Assume the IDS of $H_0$ is H\"older continuous with exponent $\delta>0$ in some open interval $\Delta_0\subset ]-\infty,E_0[$, then there exists a  constant $K_W$ depending on $d$, $E_0$, $\delta_{\pm}$,$C_{\pm}$, such that for any $\lambda\le 1$, $\Delta\subset \Delta_0$ compact, $|\Delta|$ small enough, and any $0<\gamma<1$,
\beq\label{Wegnerbis}
\E \set{\tr E_{H^{(\Lambda)}_{\lambda,\omega}}(\Delta)} \le K_W \max\left(|\Delta|^{\delta\gamma}, |\Delta|^{-\gamma {m(d)}}Q_{\mu^{(m(d))}}(|\Delta|)\right) |\Lambda|.
\eeq
In particular, if $Q_{\mu^{(m(d))}}(\eps)\leq C\eps^{\zeta}$, for some $\zeta \in ]0,1]$, then
\beq \label{Wegnerbis2}
\E \set{\tr E_{H^{(\Lambda)}_{\lambda,\omega}}(\Delta)}\leq K_W |\Delta|^{\frac{\zeta\delta}{\delta +{m(d)}}}|\Lambda|.
\eeq
\item[(c)] Assume $E\in\Delta_0\subset(\R\setminus\sigma(H_0))\cap ]-\infty,E_0[$, $\Delta_0$ compact, then there exists a  constant $K_W$, depending on $d$, $E_0$, $\delta_{\pm}$, $C_{\pm}$ and $\Delta_0$, such that for any $\lambda\le 1$ and any $\Delta\subset\Delta_0$ centered at $E$, $|\Delta|$  small enough,
\beq\label{Wegnerter}
\E \set{\tr E_{H^{(\Lambda)}_{\lambda,\omega}}(\Delta)} \le K_W  \lambda Q_{\mu^{(m(d))}}(|\Delta|)) |\Lambda|.
\eeq
\end{itemize}
\end{theorem}

We adapt the proof of \cite{CHK2}, using the basic spectral averaging estimate proved in \cite{CHK2}:
Let $H_0$ and $W$
be   self-adjoint operators on a Hilbert space $\H$, with $W\ge 0$ bounded.   Let $H_s:= H_0  + s W$ for $s \in \R$.
Then,  given $\vphi\in \H$ with
$\norm{\vphi}=1$,  for  all Borel measures $\nu$ on $\R$ and  all bounded
intervals $I\subset \R$ we have (\cite[Corollary~4.2]{CH}, \cite[Eq.~(3.16)]{CHK2}\footnote{There  the estimate  \eqref{sa}  is stated with $W$ instead of $\sqrt{W}$, with the additional hypothesis that $W\le 1$.  But a careful reading of their proof shows that they actually prove  \eqref{sa} as stated here.}
\begin{equation}\label{sa}
\int \di \nu(s)\,
\langle \vphi,\sqrt{W} \chi_I(H_s) \sqrt{W}\vphi\rangle \le Q_\nu(\abs{I}).
\end{equation}
The result is stated in \cite{CHK2} for a probability measure $\nu$ with compact support, but their proof works for an  arbitrary Borel measure $\nu$.  In particular, for $H_\omega$ as in Theorem~\ref{thmWegner}, we get, for any $\phi\in\L^2(\R^d)$,  $j\in\Z^d$, $\alpha>0$,  and any interval $I_\eps$ of length $\eps>0$,
\begin{align}
\E \lbrace|\omega_{j}|^{\alpha}\langle\phi,\sqrt{u_{j}}E_{H^{(\Lambda)}_{\lambda,\omega}}(I_\eps)\sqrt{u_{j}}\phi\rangle\rbrace  \leq  \tfrac 1 \lambda Q_{\mu^{(\alpha)}}(\epsilon)\Vert\phi\Vert^{2}.
\end{align}
As a consequence, for any trace class operator $S\ge 0$,
\begin{align}\label{lemsa}
\E \left\{ |\omega_{j}|^{\alpha} \tr \set{\sqrt{u_{j}}E_{H^{(\Lambda)}_{\lambda,\omega}}(I_\eps)\sqrt{u_{j}}S}\right\}  \leq   \tfrac 1 \lambda (\tr S) Q_{\mu^{(\alpha)}}(\epsilon).
\end{align}

\begin{proof}[Proof of Theorem~\ref{thmWegner}]
Recall that  $H_{\lambda,\omega}= H_0 + \lambda V_\omega$, $\lambda\in]0,1]$, and to alleviate notations we write $E_{\Lambda}(\Delta):=E_{H^{(\Lambda)}_{\lambda,\omega}}(\Delta)$ and $E_{0}^{\Lambda}(\Delta):=E_{H^{(\Lambda)}_{0,\omega}}(\Delta)$. To simplify the exposition we assume that the support of $u$ is smaller than the unit cube; if not the case, the proof can be modified in a straightforward way, as in \cite{CHK2}. In particular, $u_iu_j=0$ if $i\neq j$. We also introduce $\chi$ to be the characteristic function of a cube containing the support of $u$, contained in the unit cube, such that  $\chi_i \chi_j=0$ if $i\neq j$, where $\chi_j(x)=\chi(x-j)$.   With $\Delta\subset\tilde{\Delta}$, and denoting $d_{\Delta}=\mathrm{dist}(\Delta , \tilde{\Delta}^{c})$, we get
\beq
\tr(E_{\Lambda}(\Delta))=\tr(E_{\Lambda}(\Delta)E_{0}^{\Lambda}(\tilde{\Delta}))+\tr(E_{\Lambda}(\Delta)E_{0}^{\Lambda}(\tilde{\Delta}^{c})) \label{decomp1}.
\eeq

We first consider the term $\tr(E_{\Lambda}(\Delta)E_{0}^{\Lambda}(\tilde{\Delta}^{c}))$ and take care of the unboundedness of the random variables. We have,
\begin{align}
\tr(E_{\Lambda}(\Delta)E_{0}^{\Lambda}(\tilde{\Delta}^{c})) & \leq C_d(\Delta)\lambda^2\sum_{i,j\in \Lambda}\vert\omega_{i}\omega_{j}\vert \vert \tr( {u_j}E_{\Lambda}(\Delta){u_i} K_{ij})\vert \\
&  \leq C_d(\Delta)\lambda^2\sum_{i,j\in \Lambda, i\neq j}\vert\omega_{i}\omega_{j}\vert \vert \tr( {u_j}E_{\Lambda}(\Delta){u_i} K_{ij})\vert \\
& + C_d(\Delta)\lambda^2\sum_{i\in\Lambda}\vert\omega_{i}\vert^{2} \vert \tr( {u_i}E_{\Lambda}(\Delta){u_i} K_{ii})\vert
\end{align}
where
\beq
K_{ij} =  \chi_{i}(H_{0}^{\Lambda}+M)^{-2}\chi_{j},
\eeq
and
\begin{equation}
\left\Vert \left(\dfrac{H_{0}^{\Lambda}+M}{H_{0}^{\Lambda}-E_m}\right)^{2}E_{0}^{\Lambda}(\tilde\Delta^{c})\right\Vert
\leq\left(1+\dfrac{2(M+\Delta_{+})}{d_{\Delta}}+\dfrac{(M+\Delta_{+})^{2}}{d_{\Delta}^{2}}\right)=C_d(\Delta)
\end{equation}
for some $M<\infty$ such that $H_0+M\ge 1$, for example $M=1$ is enough, and where the $\chi_{i}, \forall i\in\Z^{d}$ are compactly supported functions, with support slightly larger than the $u_{i}$'s one such that $\chi_{i}u_{i}=u_{i}$. Note that $K_{ij}$ is trace class as soon as $i\neq j$ (since we assume $\supp u_j \subset \Lambda_1(j)$), as can be seen by a successive use of the resolvent identity, and by Combes-Thomas its trace class norm satisfies $\|K_{ij}\|_1 \le C_d \e^{-|i-j|}$, for $i\neq j$. It follows, as in \cite[Eqs~(4.1)-(4.4)]{CGK}, that
\begin{align}
&\sum_{i\neq j}\vert\omega_{i}\omega_{j}\vert  \left|  \tr( {u_j} E_{\Lambda}(\Delta){u_i} K_{ij})\right| \\
& \le \sum_{i\neq j}  \frac12\left(\vert\omega_{i}\vert^2\tr( {u_i}E_{\Lambda}(\Delta){u_i} |K_{ij}|)+\vert\omega_{j}\vert^2\tr( {u_j}E_{\Lambda}(\Delta){u_j} |K_{ij}^\ast|)\right) \\
& = \sum_i \vert\omega_{i}\vert^2 \left| \tr( {u_i}E_{\Lambda}(\Delta){u_i} S_i)\right|,
\end{align}
where
\beq
S_j=\frac12\sum_{i\neq j} (|K_{ij}| + |K_{ji}^\ast|)\ge 0,
\eeq
with
\beq\label{uniftracesum2}
\max_{j\in \Lambda } \tr S_j \le Q_2<\infty.
\eeq
It remains to consider the diagonal term $i=j$, that is $|\omega_i|^2 \tr( {u_i} E_{\Lambda}(\Delta){u_i} K_{ii})$. Note that $K_{ii}$ is trace class in dimension $d=1,2,3$ but not higher. To deal with the general case of arbitrary dimension we proceed as in \cite{CHK2} and perform successive Cauchy-Schwartz inequalities, getting, for any integer $m\ge 1$, for some constant $K_{d,m}<\infty$,
\begin{align}
&C_d(\Delta)|\omega_i|^2 \tr( {u_i} E_{\Lambda}(\Delta){u_i} K_{ii}) \\
& \le
\frac14 \tr( {u_i} E_{\Lambda}(\Delta) {u_i}) + K_{d,m}(C_d(\Delta) |\omega_i|)^{2^{m}} \tr( {u_i} E_{\Lambda}(\Delta){u_i} K_{ii}^{2^{m-1}}).
\end{align}
We chose $m$ so that $K_{ii}^{2^{m-1}}$ is trace class, that is, we take
$m(d):=2^{m+1}> d$, i.e., $m=[\log d / \log 2]$, where $[x]$ stands for the integer part of $x$. It follows that, using $\sum_j u_j \le 1$, uniformly in $\lambda\le 1$,
\begin{align}
& \tr(E_{\Lambda}(\Delta)E_{0}^{\Lambda}(\tilde{\Delta}^{c}))\\
& \leq \frac14 \sum_i\tr( u_i E_{\Lambda}(\Delta)u_i) + K_{d,m(d)}\lambda^2\sum_i (C_d(\Delta)|\omega_i|)^{{m(d)}} \tr( {u_i} E_{\Lambda}(\Delta){u_i} \tilde{S_i}) \\
& \le
\frac14 \tr E_{\Lambda}(\Delta) + K_{d,m(d)}^{'}\lambda^2\sum_i \left(\frac{|\omega_i|}{d_\Delta}\right)^{{m(d)}} \tr( {u_i} E_{\Lambda}(\Delta){u_i} \tilde{S_i}) ,
\end{align}
where
\beq
\tilde{S_i}=S_i + K_{ii}^{2^{m(d)-1}}\ge 0,
\eeq
is a trace class operator. We apply \eq{lemsa} to finish the bound:
\beq
\E\tr(E_{\Lambda}(\Delta)E_{0}^{\Lambda}(\tilde{\Delta}^{c}))\leq \frac14 \E\tr E_{\Lambda}(\Delta) + C_{d}^\prime\lambda\frac{Q_{\mu^{(m(d))}}}{d_{\Delta}^{{m(d)}}}(\vert\Delta\vert)\vert\Lambda\vert. \label{control12}
\eeq
We now turn to the first term of the right hand side in \eqref{decomp1}, that is $\tr(E_{\Lambda}(\Delta)E_{0}^{\Lambda}(\tilde{\Delta}))$. To get the general Wegner estimate \eqref{Wegner} the latter is treated as in \cite{CHK2}, using either the unique continuation principle for the free Hamiltonian, or, in the Landau case, explicit properties of the Landau Hamiltonian. Note that we then incorporate $d_\Delta$ in the constant.
To get \eqref{Wegnerbis}, we control $\tr(E_{\Lambda}(\Delta)E_{0}^{\Lambda}(\tilde{\Delta}))$ using  the hypothesis on the IDS of $H_0$, that is $\tr E_{H_0}(\tilde{\Delta})\le C |\tilde{\Delta}|^\delta |\Lambda|$. In this case, we need $d_{\Delta}$ to be small enough and it then remains to control the growth of the constant in the second term of the r.h.s. of \eqref{control12}.
Taking $d_{\Delta}=\eps^{\gamma}$, with $0<\gamma<1$, and using $Q_{\mu^{(m(d))}}(\vert\Delta\vert)\leq C\eps^{\zeta}$ if $\mu$ is $\zeta$-H\"older continuous, we get, with a new constant $K_W$, and $\eps$ small enough so that $\tilde{\Delta}\subset\Delta_0$,
\begin{align}
\E {\tr E_{\Lambda}(\Delta)} & \leq K_W \mathrm{max}\left(\eps^{\gamma\delta},\frac{1}{\eps^{{m(d)}\gamma}}Q_{\mu^{(m(d))}}(\eps)\right)\vert\Lambda\vert \\
& \leq K_W \mathrm{max}\left(\eps^{\gamma\delta},\eps^{\zeta-{m(d)}\gamma}\right)\vert\Lambda\vert \\
& \leq K_W \eps^{\frac{\zeta\delta}{\delta +{m(d)}}}|\Lambda|
\end{align}
where we have chosen $\gamma$ such that $\gamma\delta=\zeta-{m(d)}\gamma$.

Finally, in the particular case of \eqref{Wegnerter}, $\tr E_{H_0}(\tilde{\Delta})=0$ as long as $\tilde{\Delta}\subset \Delta_0$.
\end{proof}

The following theorem contains an extension of \cite{HKS} to unbounded random variables. We set, for $E\in\R$, $P_{\lambda,E ,\omega}= \chi_{]-\infty,E]}(H_{\lambda,\omega})$, the Fermi projection.

\begin{theorem}\label{holderc} Consider $H_{\lambda,\omega}$  with $0< \lambda\le 1$
Assume that the IDS of $H_0$ is H\"older continuous in $E\in \Delta_0$ an open interval. Then for some $\nu>0$ and $C_{\Delta_0}<\infty$, for any $E,E'\in \Delta_0$, $|E-E'|$ small enough, we have uniformly in $0\le \lambda\le 1$,
\beq\label{IDS2}
\max_{u\in\Z^2}\E\set{\norm{\chi_0\left(
P_{\lambda, E ,\omega} -
P_{\lambda, E',\omega}\right) \chi_u }_1 }
\le  C_{\Delta_0} |E-E'|^\nu,
\eeq
and for some $\nu^\prime>0$,  for all $E\in \Delta_0$, for all $\lambda^\prime,\lambda^{\prime\prime}\in[0,1]$, $|\lambda^{\prime\prime}-\lambda^\prime|$ small enough,
\beq\label{IDSlambda2}
\max_{u\in\Z^2} \E\set{\norm{\chi_0\left(
P_{\lambda^{\prime},E ,\omega} -
P_{\lambda^{\prime\prime} , E ,\omega}\right) \chi_u }_1 }
\le  C_{\Delta_0} |\lambda^{\prime\prime}-\lambda^\prime|^{\nu^\prime}.
\eeq

\end{theorem}

\begin{proof}
Eq.~\eqref{IDS2} follows from Cauchy-Schwarz and the continuity of the Integrated Density of States of $H_{\lambda,\omega}$ given by Theorem~\ref{thmWegner} Eq.~\eqref{Wegnerbis2}. We turn to \eqref{IDSlambda2}.
Let $ E \in \Delta_0$ and
$ \lambda^\prime,  \lambda^{\prime\prime}  \in   [\lambda_1,\lambda_2]$ possibly containing 0.
We let   $\gamma=  |\lambda^\prime - \lambda^{\prime\prime} |^{\alpha}$,
where $\alpha \in (0,1)$ will be chosen later. Let $f(t)$ be a smooth decaying switch function, equal to $1$ for $t\le 0$ and $0$ for $t\ge 1$.
We set $g(t) = f\left( \frac { t- (E-\gamma)} {\gamma}\right)$; note
$ g \in C^\infty(\R)$, with $0 \le g(t) \le 1$,
$g(t)=1$ if $ t \le
E - \gamma$,
$g(t) =0$ if $t\ge E$. We write
\begin{align}\label{Pg}
& P_{\lambda ^\prime,E ,\omega}-  P_{\lambda ^{\prime\prime},E,\omega}=
\left\{ P_{\lambda ^\prime,E ,\omega}-g^2(H_{\lambda ^\prime,\omega})\right\}\\
& \qquad  \qquad  \qquad
+ \left\{g^2(H_{\lambda^\prime,\omega}) -
g^2(H_{\lambda^{\prime\prime},\omega})\right\} +
\left\{g^2(H_{\lambda^{\prime\prime},\omega}) -
P_{\lambda^{\prime\prime},E,\omega}\right\}\notag .
\end{align}
By construction, for any $\lambda \ge 0$ we have
\begin{equation}
0 \le P_{\lambda,E ,\omega} -
g^2(H_{\lambda ,\omega}) \le
P_{\lambda ,E ,\omega} - 
P_{\lambda , E - \gamma,\omega}\, ,
\end{equation}
and thus, for  $\lambda^\#= \lambda^\prime, \lambda^{\prime\prime}  $
and any  $u \in \Z^2$,
we have
\begin{align} \label{bounduseW}
&\qnorm{ \chi_0\left(
P_{\lambda ^{\#},E ,\omega} -
g^2(H_{\lambda ^\#,\omega}) \right) \chi_u}_1\\
& \notag \ \le \qnorm{ \chi_0\left(
P_{\lambda ^{\#},E ,\omega} -
g^2(H_{\lambda ^\#,\omega}) \right)^{\frac 12} }_2
\qnorm{ \left(
P_{\lambda ^{\#},E ,\omega} -
g^2(H_{\lambda ^\#,\omega})  \right)^{\frac 12}  \chi_u
}_2\\
&\ \notag  =\qnorm{\chi_0\left(
P_{\lambda ^{\#},E ,\omega} -
g^2(H_{\lambda ^\#,\omega}) \right) \chi_0 }_1\\
& \ \notag \le \qnorm{\chi_0\left(
P_{\lambda ^{\#},E ,\omega} -
P_{\lambda ^{\#}, E - \gamma,\omega}\right) \chi_0 }_1
\le  C_{\Delta_0} \gamma^\nu.
\end{align}
To control the middle term in the r.h.s. of \eqref{Pg}, we proceed as in \cite[Eq.~(3.8)]{HKS} and sequel. In the Helffer-S\"ojstrand formula, one needs to go to the $(4+2d)$th order.  The term corresponding to \cite[Eq.~(3.15)]{HKS} is controled as follows (we denote by $R_{\lambda,\omega}(z)$ the resolvent of $H_{\lambda,\omega}$):
\begin{align}
& \|R_{\lambda,\omega}(z) V_\omega R_{\lambda^\prime,\omega}(z) V_\omega R_{\lambda,\omega}(z) \chi_0\|
\\
& \le
\sum_{j,k\in\Z^d} |\omega_j\omega_k|\|R_{\lambda,\omega}(z) u_j R_{\lambda^\prime,\omega}(z) u_k R_{\lambda,\omega}(z) \chi_0\| \\
& \le
\sum_{j,k\in\Z^d} |\omega_j\omega_k| |\Im z|^{-3} \e^{-c|\Im z||j-_k|} \e^{-c|\Im z||k|} .
\end{align}
It follows, using the Combes-Thomas inequality, that
\begin{align}
& \E \| \chi_u g(H_{\lambda,\omega})\|_1 \|R_{\lambda,\omega}(z) V_\omega R_{\lambda^\prime,\omega}(z) V_\omega R_{\lambda,\omega}(z) \chi_0\|
\\
& \le
\sum_{j,k\in\Z^d} \left(\E |\omega_j\omega_k| \| \chi_u g(H_{\lambda,\omega})\|_1 \right)  |\Im z|^{-3} \e^{-c|\Im z||j-_k|} \e^{-c|\Im z||k|} \\
& \le
C(I,d) |\Im z|^{-3} \sum_{j,k\in\Z^d} \e^{-c|\Im z||j-_k|} \e^{-c|\Im z||k|}
\\
& \le C(I,d) |\Im z|^{-3-2d},
\end{align}
by Theorem~\ref{thmSGEE}. The term corresponding to \cite[Eqs.~(3.16)-(3.18)]{HKS} is controled in a similar way using Theorem~\ref{thmSGEE}.
\end{proof}

\end{document}